\documentclass[10pt,a4paper]{extarticle}
\pdfoutput=1

\usepackage[T1]{fontenc}
\usepackage[utf8]{inputenc}   
\usepackage[pdftex]{graphicx}                        
\usepackage[paper=a4paper, tmargin=1.75cm, bmargin=1.75cm, lmargin=1.25cm, rmargin=1.25cm, centering]{geometry}
\usepackage{mathpazo}
\usepackage{inconsolata}                             
\usepackage[hyphens,obeyspaces,spaces]{url}
\usepackage[svgnames]{xcolor}
\usepackage{multicol}
\usepackage{amsmath}
\usepackage{amsfonts}
\usepackage{amssymb}
\usepackage{amsbsy}
\usepackage{setspace}
\usepackage{xspace}                                
\usepackage[detect-all]{siunitx}
\usepackage[font=small]{caption}
\usepackage{subcaption}
\usepackage{bm}
\usepackage{float}                                 

\graphicspath{{figures/}}                           
\newcommand{\subrm}[1]{\ensuremath{_{\text{#1}}}}

\providecommand{\degree}{\ensuremath{^\circ}\xspace}    
\renewcommand{\degree}{\ensuremath{^\circ}\xspace}      
\newcommand{\silent}[1]{}
\newcommand{\figref}[1]{Fig.~\ref{#1}}
\newcommand{\vv}[1]{\bm{\mathrm{#1}}}                   

\newcommand{\norm}[1]{\left\lVert#1\right\rVert}

\makeatletter      
   \newcommand{\figcaption}{\captionsetup{type=figure} \def\@captype{figure}\caption}            
   \renewcommand{\@biblabel}[1]{#1.}    
\makeatother

\begin{document}

\begin{center}
   {\huge\bf The geometric phase of rotations and 3D coordinate transformations} \\[18pt] 
   \textbf{Luis Garza-Soto}${}^{1,2*}$ and \textbf{Nathan Hagen}${}^2$ \\[8pt]
   ${}^1$Department of Aerospace Structures and Materials, Delft University of Technology, Kluyverweg 1, 2629 HS, Delft, The Netherlands \\
   ${}^2$Utsunomiya University, Dept of Optical Engineering, 7-2-1 Yoto, Utsunomiya, Tochigi, Japan 321-8585 \\
   ${}^{*}$email: L.A.GarzaSoto@tudelft.nl
\end{center}

\vspace{1mm}
\setlength\columnsep{15pt}

\begin{abstract}
   {\noindent}The wave description of geometric phase uses the superposition of light waves to explain the geometric phase's origin. While our previous work focused on a basis of linearly polarized waves, here we show that the same concepts can be applied to circularly polarized waves, and to any case in which a rotator is itself subjected to rotation. As with a linear polarization basis, we show that the addition of two vectors (rotators) with different orientations and magnitudes causes the orientation of the resulting vector to shift towards the component vector of greater magnitude, i.e. it introduces a geometric phase. We illustrate this approach with two classic examples of the geometric phase of rotations in space: a system of three fold mirrors, and the helical coiled fiber. In both cases we show that it is possible to derive the phase shift directly from the electromagnetic wave vector without needing to resort to mathematical abstractions such as differential geometry, or calculating solid angles in the space of directions.
\end{abstract}

\begin{multicols}{2}

\section{Introduction}

Geometric phases are often separated into different types. In this classification, Pancharatnam-Berry (PB) phase is one type, used to describe phase shifts resulting from transformations of the polarization state~\cite{Gutierrez-Vega2011,Garza-Soto2023a}. The spin-redirection phase, or Rytov-Vladimirsky-Berry (RVB) phase, describes phase shifts produced by transforming a wave's propagation direction~\cite{Rytov1938,Vladimirski1941,Berry1987b,Tavrov1999}. A third common type, used for quantum systems, is generally described just as Berry phase~\cite{Berry1984,Aharonov1987}. In mechanical systems, the geometric phase has been called the Hannay angle, for situations such as when a fast spinning object is itself subjected to an additional rotation, much like a spinning planet orbiting the Sun~\cite{Hannay1985,Berry2011}. 

While the various geometric phases are often treated separately, we show in the discussion below that the recently developed superposition model of geometric phase~\cite{garza2023wave} can deal with PB phase, RVB phase, and the Hannay angle under the same formalism, using only elementary mathematics and visualizable models of the phase. This visualizability adds insight by clarifying how geometric phase arises from coordinate transformations and wave superposition. Although the gauge dependence of geometric phase has often been mentioned as a problem for non-cyclical paths, we demonstrate that it is instead an essential feature of the phenomenon and is readily calculated.

The first attempts to measure RVB phase experimentally were done with a helically twisted fiber, such that the propagation direction vector of the electromagnetic wave inscribes a conical domain in the space of propagation directions. The geometric phase $\gamma$ is related to the solid angle $\Omega$ of the domain traced on the sphere of directions via $\gamma = -\Omega$ (with an implied change in units from steradians to radians), allowing us to calculate the phase shift induced from the propagation path geometry. In the typical analysis, using a polarization basis of right-circular and left-circular states, $\gamma$ shows up as a phase delay between the two basis states. If, on the other hand, the state of polarization is represented using a linear polarization basis, then $\gamma$ appears not as a phase delay but rather as a rotation of the polarization angle. This difference with how $\gamma$ manifests depending on the choice of basis is widely known but it is not well-known how to connect the two perspectives under one formalism. Sections~\ref{sec:vectors}\,\&\,\ref{sec:coupled_rotations} try to bridge this gap by representing states using rotating vectors (phasors) and demonstrate how phases shift when we sum two or more phasors --- equivalent to adding two or more circularly polarized waves of different amplitudes. 

\begin{figure*}
    \centering
    \includegraphics[width=\linewidth]{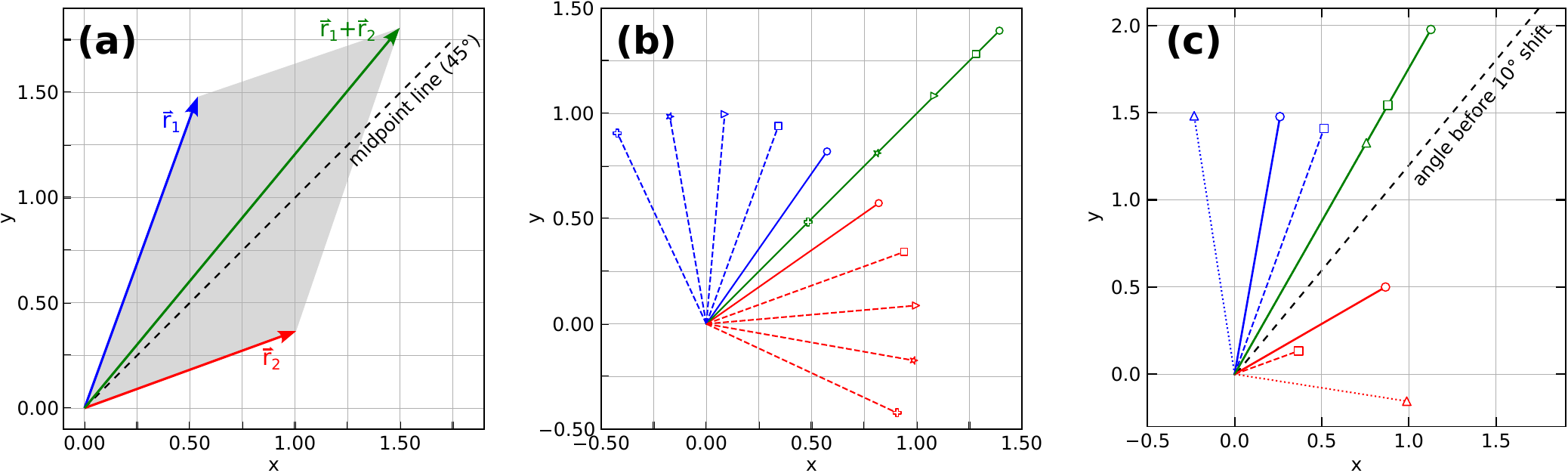}
    \caption{(a) Adding vectors with parameters $A_1 = 1.5$, $\theta_1 = 70\degree$; $A_2 = 1$, $\theta_2 = 20\degree$. (b) Adding vectors with parameters $A_1 = 1$, $\theta_1 = 45\degree + \Delta$; $A_2 = 1$, $\theta_2 = 45\degree - \Delta$; where $\Delta \in \{ 10\degree, 25\degree, 40\degree, 55\degree, 70\degree\}$. (c, solid lines) Adding vectors with $A_1 = 1.5$, $A_2 = 1$, $\bar{\theta} = 45\degree$, $\phi = 10\degree$, $\delta = 50\degree$; (c, blue dashed lines) $A_1 = 1.5$, $A_2 = 0.39$, $\bar{\theta} = 45\degree$, $\phi = 0\degree$, $\delta = 50\degree$; (c, red dotted lines) $A_1 = 1.5$, $B = 1$, $\bar{\theta} = 45\degree$, $\phi = 0\degree$, $\delta = 107.8\degree$. Note that the solid lines in (c) are the same as the vectors drawn in (a) but rotated by $+10\degree$.}
    \label{fig:adding_vectors3}
\end{figure*}

This result therefore generalizes our previous work, which focused on superpositions of linear polarization basis states, and only implicitly considered circular polarization basis states or systems that use rotating vectors for a basis. Where changes of phase in a linear polarization state induce modulations in the amplitude (i.e. $\cos (kz)$), changes of phase in a circular polarization state are coupled to rotations of the polarization vector (i.e. $e^{i \theta}$). Representing the same polarization state in each basis consistently requires taking some care. As a result, Secs~\ref{sec:vectors}\,\&\,\ref{sec:coupled_rotations} go into detail about how rotating vectors can be treated with the same phase formalism as can linear states.

Our generalization also extends beyond the analysis of geometric phase in light. Considering that any oscillation (mechanical, electrical, etc.) is a wave, the superposition model of geometric phase is able to make predictions for any system of combined oscillations. For example, we can analyze the elliptical rotation of a planet orbiting the Sun in terms of circular components, using deferents and epicycles just as Ptolemy and Copernicus did, and calculate a geometric phase with respect to a chosen reference. However, we only consider cases in which the two component oscillations have the same rotation frequency.

Finally, we show that our approach provides a natural method for treating two geometric phases that are often considered separately --- Pancharatnam-Berry phase and Rytov-Vladimirsky-Berry phase --- under one framework. Since all experimental demonstrations of RVB phase that we are aware of have analyzed geometric phase using a circular polarization basis, we demonstrate here that the superposition model of geometric phase has no difficulty doing the same, despite the apparent differences in the way that phase is represented under linear and circular bases.

While the idea of combining RVB and PB phases is not new~\cite{Nityananda2014,Jiao1989,Tavrov2000}, prior work relied on using differential geometry or the solid angle rule to obtain results. These abstractions obscure the underlying physical behavior of the wave and how the result can be derived entirely from elementary methods from changes in the polarization state and changes in coordinates. The two types of phase are not linearly additive, since they depend on the coordinate basis and on the composition of the polarization state in that basis. But once we choose the gauge, we can represent the state in that basis, and the resulting geometric phase is readily obtained, telling us where we can expect to find the wave's peak. 

\section{The geometric phase of static vector addition} \label{sec:vectors}

In order to illustrate the relationship of vector addition to geometric phase, we start with a standard representation of vectors, using their length $A$ and direction $\theta$, where the direction angle is defined with respect to the $x$-axis, and counter-clockwise angles being positive. We consider the addition of vectors
\begin{equation}
\begin{aligned}
    \vv{r}_1 &= A_1 \cos(\theta_1) \hat{x} + A_1 \sin(\theta_1) \hat{y} \, , \\
    \vv{r}_2 &= A_2 \cos(\theta_2) \hat{x} + A_2 \sin(\theta_2) \hat{y} \, ,
\end{aligned}
\end{equation}
as shown graphically in \figref{fig:adding_vectors3}(a). The orientation angle $\theta_3$ of their sum is easily obtained by the harmonic addition theorem. Using the theorem's symmetric form, the orientation angle of $\vv{r}_3$ is written as the angular shift $\gamma$ with respect to the angle $\bar{\theta}$ bisecting the two components: $\theta_3 = \bar{\theta} + \gamma$, where $\bar{\theta} = (\theta_1 + \theta_2) / 2$, and $\gamma$ is given by
\begin{equation}\label{eq:gamma}
    \tan (\gamma) = \frac{A_1 \sin(\theta_1) + A_2 \sin(\theta_2)}{A_1 \cos(\theta_1) + A_2 \cos(\theta_2)} \ .
\end{equation}
This equation is equivalent to the geometric phase $\gamma$ induced by the superposition of two co-polarized electromagnetic waves~\cite{garza2023wave}, where the wave amplitudes $A_1$ and $A_2$ correspond to the magnitude of the vectors.

When the lengths of $\mathbf{r}_1$ and $\mathbf{r}_2$ are equal, the orientation of the sum vector $\vv{r}_3$ will bisect the angle between the components regardless of their individual orientation angles, as shown in \figref{fig:adding_vectors3}(b). If we use the bisecting angle as a reference, $\bar{\theta}$, instead of the $x$-axis, then the component vector orientation angles become symmetrical about the reference, $\theta_1 = \bar{\theta} + \delta/2$, $\theta_2 = \bar{\theta} - \delta/2$, and the orientation angle shift due to summing the two takes on the simplified expression~\cite{garza2023wave}
\begin{equation}\label{eq:symmetricGphase}
    \tan (\gamma) = \tan \Big( \frac{\delta}{2} \Big) \frac{A_1 - A_2}{A_1 + A_2} \ ,
\end{equation}
where $\delta$ is the angle between $\mathbf{r}_1$ and $\mathbf{r}_2$.

This vector representation is useful as a step towards the further developments we show in Sec.~\ref{sec:coupled_rotations}. However, it also provides another point of view for observing geometric phase behavior. A change in the angle of the phasor $\vv{r}_3$ by an amount $\phi$ can be achieved by rotating both input components $\vv{r}_1$ and $\vv{r}_2$ by the same angle, as shown in \figref{fig:adding_vectors3}(c) for the case of $\phi = 10\degree$. This is equivalent to adding a propagation phase $\phi$ to an electromagnetic wave. If we compare the result to \figref{fig:adding_vectors3}(a), we see that the overall phase of the resulting wave, $\theta_3$, has of course shifted exactly by $\phi$. However, we can obtain the same phase shift through a geometric phase instead. One way would be to keep the original orientation of the components (i.e., $\phi = 0$) and change only their relative lengths, as in the dashed vectors of \figref{fig:adding_vectors3}(c). In the example of \figref{fig:adding_vectors3}(c), this is accomplished by changing the length of $A_2$ from 1 to 0.39, giving a shift of $\gamma = 10\degree$ with respect to the original input of $A_2 = 1$.

Another way to obtain the same geometric phase shift is to keep the original magnitudes but change the relative angle $\delta$ between them, as in the dotted vectors of \figref{fig:adding_vectors3}(c). Since the two input vectors are not the same length, changing the separation angle $\delta$ causes a phase shift. In this case, changing $\delta$ from 45\degree to 107.8\degree creates a shift of 10\degree from the wave's starting position.

\section{The geometric phase from adding rotating vectors}\label{sec:coupled_rotations}

A circular rotation is represented by a vector of constant magnitude whose orientation angle $\theta = \omega t$ is governed by a rotation rate $\omega$ and time $t$. The position of a point on a spinning disc relative to its center and the electric field of a circularly polarized beam of light are examples. For circular rotation, the vector can be written as
\begin{equation}
   \mathbf{r} = A \cos(\omega t + \phi) \hat{x} + A \sin(\omega t + \phi) \hat{y} \, ,
\end{equation}
where an increasing orientation angle is a counter-clockwise (CCW) rotation, and $\phi$ is the initial angle. While this shows a rotating vector in a fixed frame, it is also possible to write $\vv{r}$ in terms of a rotating frame:
\begin{equation}
   \mathbf{r}\subrm{rot} = A \cos(\phi) \hat{x}' + A \sin(\phi) \hat{y}' \, ,
\end{equation}
where $\hat{x}'$ \& $\hat{y}'$ are the Cartesian coordinates in a frame rotating at frequency $\omega$ with starting phase $\phi_0 = 0$.

For the case of elliptical rotations (corresponding to the electric field vector of elliptically polarized light), the $x$- and $y$-component amplitudes and phases will differ. This form shows up below when we consider the case of adding two counter-rotating vectors.

\subsection{Addition of rotations with the same sense of rotation}

We consider the addition of two circular CCW rotations that are $\pm \frac{1}{2}\delta$ phase apart. In a fixed Cartesian basis,
\begin{equation}\label{eq:CCWrotations}
\begin{aligned}
   \mathbf{r}_1 &= A_1 \cos(\omega t + \tfrac{1}{2} \delta + \phi_0)\hat{x} + A_1 \sin(\omega t + \tfrac{1}{2} \delta + \phi_0)\hat{y} \, , \\
   \mathbf{r}_2 &= A_2 \cos(\omega t - \tfrac{1}{2} \delta + \phi_0)\hat{x} + A_2 \sin(\omega t - \tfrac{1}{2} \delta + \phi_0)\hat{y} \, ,
\end{aligned}
\end{equation}
for starting phase $\phi_0$. The sum of these two is given by
\begin{equation}
   \vv{r}_3 = A_3 \cos [\bar{\theta} (t) + \gamma] \hat{x} + A_3 \sin [\bar{\theta} (t) + \gamma] \hat{y} \, ,
\end{equation} 
for $\bar{\theta}= \omega t + \phi_0$ the bisecting angle between $\mathbf{r}_1$ and $\mathbf{r}_2$, $\gamma$ the geometric phase shift given by \eqref{eq:symmetricGphase}, and
\begin{equation}
    A_3 = \sqrt{A_1^2 + A_2^2 + 2 A_1 A_2 \cos(\delta)} \, .
\end{equation}
When the phase difference $\delta$ between the two vectors is zero, their sum is simply
\begin{equation}\label{eq:r3}
   \mathbf{r}_3 = A_3 \cos(\omega t + \phi_0) \hat{x} + A_3 \sin(\omega t + \phi_0) \hat{y} \, ,
\end{equation} 
which traces a circle of radius $A_3 = A_1 + A_2$. Figure~\ref{fig:RotationsDelta0} shows the three circles traced by the two input vectors individually, and by their sum vector $\vv{r}_3$. The center of the red circle is placed at the tip of $\mathbf{r}_1$, rather than at the origin, for visualizing the vector sum. Vector $\vv{r}_3$ is indicated by the green arrow in \figref{fig:RotationsDelta0}. The bisecting angle $\bar{\theta}$ between $\mathbf{r}_1$ and $\mathbf{r}_2$ corresponds to a co-rotating reference, giving the orientation angle that each component would have at time $t$ if $\delta = 0$. For this setup, the sum vector points in the same direction as the reference ($\theta_3 = \bar{\theta}$). Note that the dot-dashed line pointing in the opposite direction of the reference arrow represents the external bisecting angle.

When superposing two rotations with the same sense of rotation and the same frequency $\omega$, the result will always be a circle for the path of $\vv{r}_3$.

When the phase difference between the two vectors is nonzero ($\delta \neq 0$), the sum vector's orientation angle shifts by $\gamma$ with respect to the reference angle $\bar{\theta}$. Figure~\ref{fig:RotationsDelta0p3pi} illustrates an example where $A_1 = 1$, $A_2 = 0.7$, $\phi_0 = 0$, and $\delta = 0.3\pi$. Again, the inputs are drawn in blue and red, while their sum is drawn in green, with their corresponding circles indicating their traces over time. In the co-rotating reference frame, the $\omega t + \phi_0$ carrier frequency and initial reference phase drop out from the modulation term arguments, leaving the stationary result
\begin{equation}
\begin{aligned}
   \mathbf{r}_1 = A_1 \cos(+\tfrac{\delta}{2}) \hat{x}' + A_1 \sin(+\tfrac{\delta}{2}) \hat{y}' \, , \\
   \mathbf{r}_2 = A_2 \cos(-\tfrac{\delta}{2}) \hat{x}' + A_2 \sin(-\tfrac{\delta}{2}) \hat{y}' \, .
\end{aligned}
\end{equation} 

\begin{center}
    \includegraphics[width=0.75\linewidth]{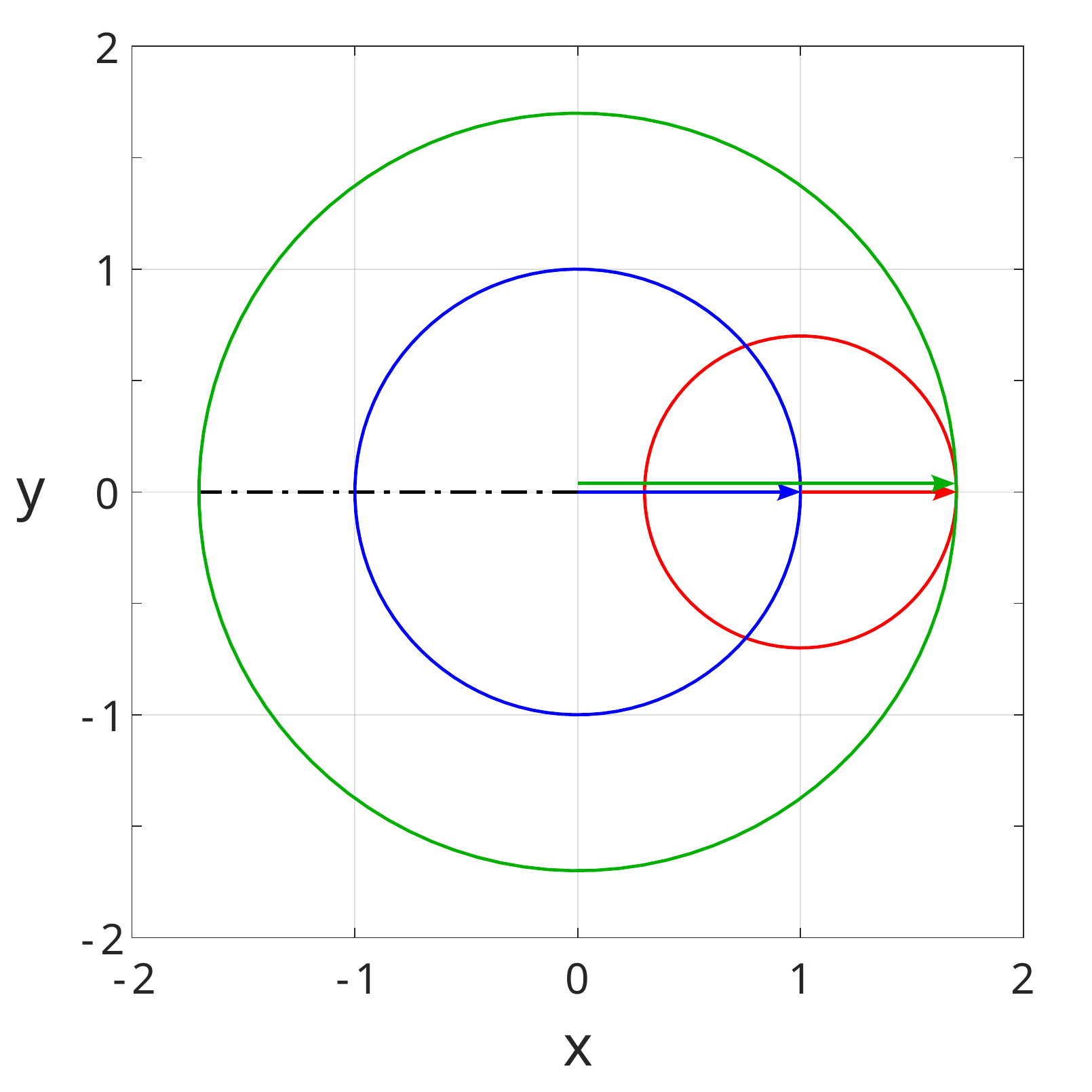}
    \figcaption{Addition of two in-phase rotating vectors (blue and red). The sum vector (green) has been shifted up for visibility. The origin of the red vector has been moved to the tip of the blur vector to better visualize their addition.}
    \label{fig:RotationsDelta0}
\end{center}

\begin{center}
    \includegraphics[width=0.75\linewidth]{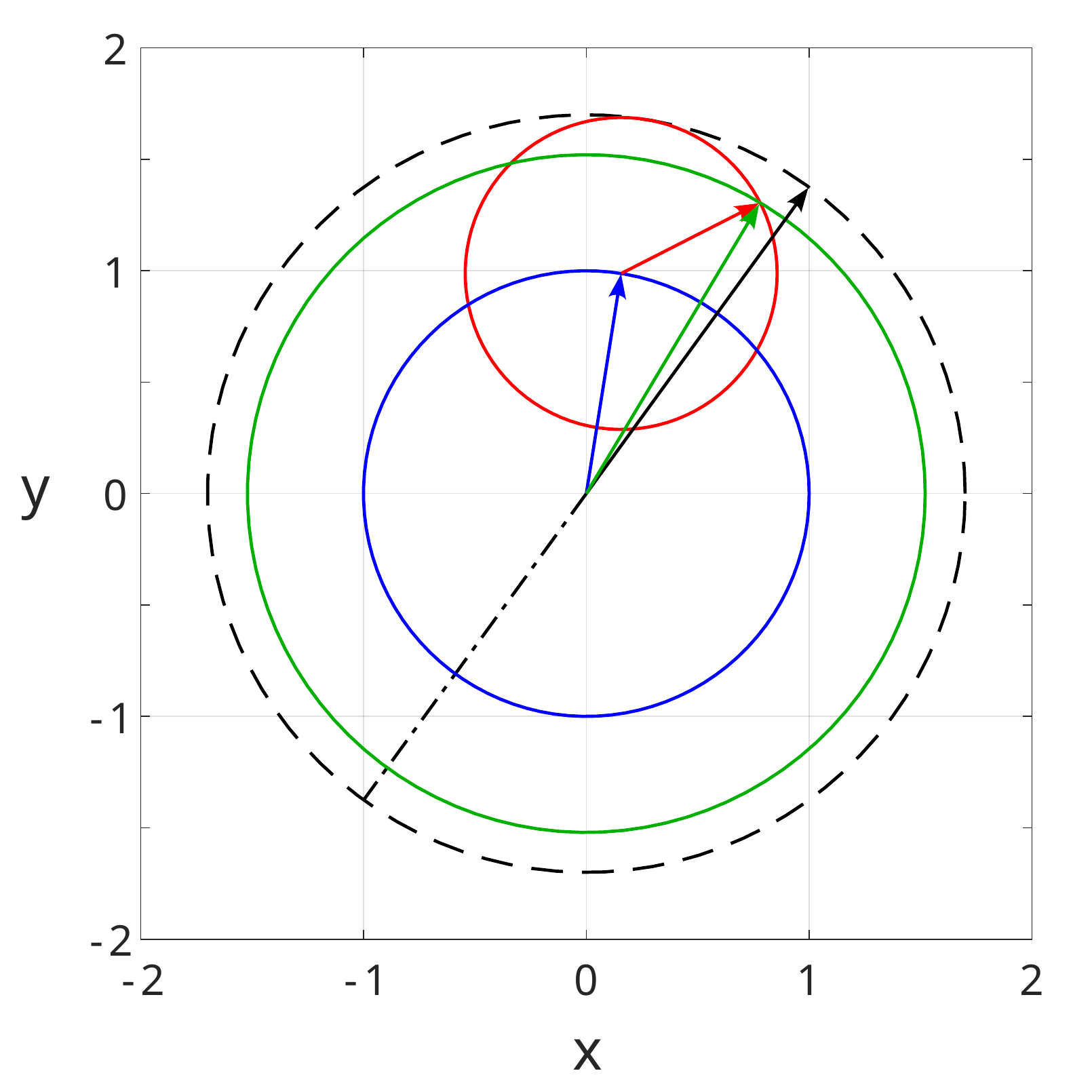}
    \figcaption{Addition of two co-rotating vectors (blue and red). Because the two input vectors are not in phase, the sum vector (green) has a smaller radius than occurs for in-phase addition (dashed black). For the case illustrated here, $A_1 = 1$, $A_2 = 0.7$, $\delta = 0.3\pi$, at time-snapshot $\omega t = 0.3\pi$. The black arrow indicates the angle bisecting the blue and red arrows, when the red arrow is placed at the origin. Visualization~1 shows an animation of this model for various values of $\delta$.}
    \label{fig:RotationsDelta0p3pi}
\end{center}

The phase shift between the reference angle $\bar{\theta}$ and the orientation of the vector sum $\theta_3$ is given by the geometric phase $\gamma = \theta_3 - \bar{\theta}$, as calculated in \eqref{eq:symmetricGphase}. In the example of \figref{fig:RotationsDelta0p3pi}, the phase difference between components is $\delta = 0.3\pi$, producing an angular shift $\gamma = 5.14 \degree$ relative to the reference angle $\bar{\theta} = 54 \degree$ at a time corresponding to $\omega t = 0.3\pi$ and initial phase angle $\phi_0 = 0$.

Whereas \figref{fig:RotationsDelta0p3pi} shows a single snapshot, Visualization~1 shows an animation of the four vectors (two input vectors, their sum vector, and the phase reference vector) for three full turns of rotation, and for different values of phase difference $\delta$. As in the case of harmonic addition of waves~\cite{garza2024differences}, the animation demonstrates that when the phase delay reaches $\delta = \pm \pi$, the phase reference suddenly shifts by $\pi$ in order to maintain its position between the two input vectors.

\subsection{Opposite sense of rotation}

Next, we consider the addition of a circular CCW rotation and a circular CW rotation. The opposite sense of the CW rotation results in a change of sign of the arguments:
\begin{equation}\label{eq:ellipse_components}
\begin{aligned}
   \mathbf{r}_1 &= A_1 \cos(\omega t + \tfrac{1}{2} \delta + \phi_0) \hat{x} + A_1 \sin(\omega t + \tfrac{1}{2} \delta + \phi_0) \hat{y} \, , \\
   \mathbf{r}_2 &= A_2 \cos(-\omega t + \tfrac{1}{2} \delta - \phi_0) \hat{x} + A_2 \sin(-\omega t + \tfrac{1}{2} \delta - \phi_0) \hat{y} \, .
\end{aligned}
\end{equation}
Note that the phase $\delta/2$ is positive for both rotations but it represents a "forward" propagation for $r_1$ and a "backwards" propagation for $r_2$, since it is opposite in sign to the time evolution of the latter. Adding the two vectors gives
\begin{equation}\label{eq:t+delta/2}
    \tan (\theta_3) = \frac{A_1 \sin(\omega t +\delta/2 + \phi_0) - A_2 \sin(\omega t -\delta/2 + \phi_0)}{A_1 \cos(\omega t + \delta/2 + \phi_0) + A_2 \cos(\omega t - \delta/2 + \phi_0)} \ ,
\end{equation}
obtained by substituting \eqref{eq:ellipse_components} into \eqref{eq:gamma} and solving. Equation~\ref{eq:t+delta/2} corresponds closely to \eqref{eq:gamma} for adding harmonic waves in 1D, but comprises a \emph{difference} of sine terms in the numerator, rather than a sum. Another difference from the case for same-sense rotations is that superposing two rotations of the same frequency $\omega$ but having the \emph{opposite} sense of rotation will in general produce an ellipse rather than a circle.

\begin{center}
    \includegraphics[width=0.75\linewidth]{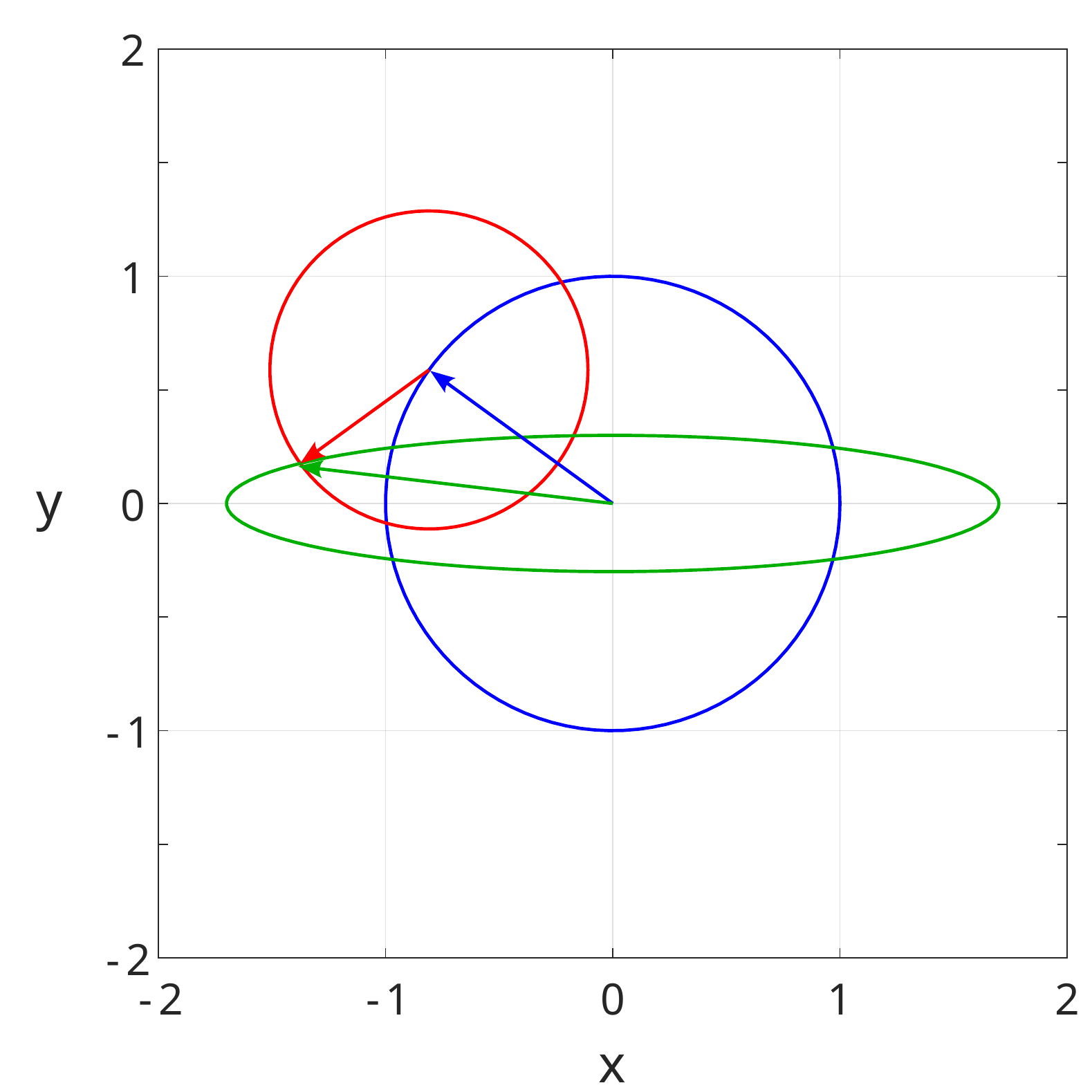}
    \figcaption{Addition of two counter-rotating vectors for $A_1 = 1$, $A_2 = 0.7$, $\delta = 0$, at time-snapshot $\omega t = 0.8\pi$. The origin of the red vector has been moved to the tip of the blur vector to better visualize their addition. Visualization~2 shows an animation of this model for $0 \leq \omega t \leq 4\pi$.}
    \label{fig:RotationsOppositeSenseDelta0}
\end{center}

\begin{figure*}
    \centering
    \includegraphics[width=1\linewidth]{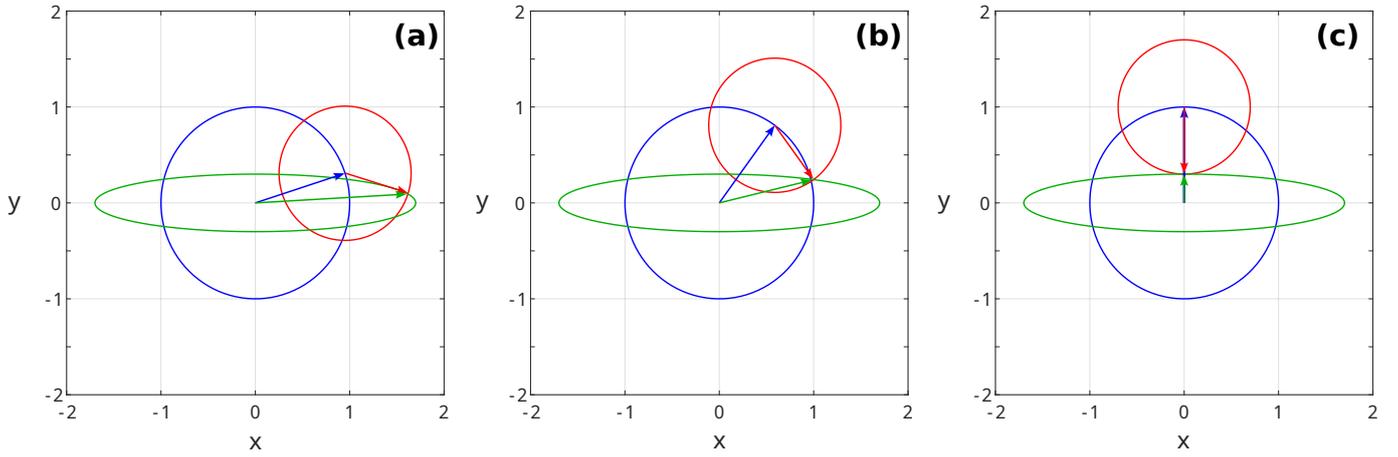}
    \caption{Addition of two counter-rotating vectors for $A_1 = 1$, $A_2 = 0.7$, and $\delta = 0$ at three time-snapshots: (a) $\omega t = 18\degree$, (b) $\omega t = 54\degree$, (c) $\omega t = 90\degree$.}
    \label{fig:RotationsOppositeVsRef}
\end{figure*}

When the two counter-rotating waves start in phase ($\delta = 0$), the sum rotation simplifies to
\begin{equation}
   \mathbf{r}_3 = (A_1 + A_2) \cos(\omega t + \phi_0) \hat{x} + (A_1 - A_2) \sin(\omega t + \phi_0) \hat{y} \, ,
\end{equation} 
a result similar to that of \eqref{eq:r3}, but the length of the $y$-component has changed from a sum to a difference. An example of this case is illustrated in \figref{fig:RotationsOppositeSenseDelta0}, for $A_1 = 1$, $A_2 = 0.7$, $\delta = 0$, at time-snapshot $\omega t = 0.8\pi$. Since $\vv{r}_1$ and $\vv{r}_2$ both share the same rotation rate $\omega$ and initial phase reference angle $\phi_0$, the angles of all three vectors advance together with time $t$, and phase difference $\delta$ remains zero.

When $\delta=0$, the angle of the sum vector $\vv{r}_3$ will be the orientation angle of the reference vector:
\begin{equation}\label{eq:ellipsePhase}
    \tan (\theta\subrm{ref}) = \tan (\omega t + \phi_0) \, \frac{A_1 - A_2}{A_1 + A_2} \ .
\end{equation}
Note that when the component rotations were both CCW (Sec.~\ref{sec:coupled_rotations}a), the co-rotating reference matched the orientation angle of the components before shift $\delta$, $\theta\subrm{ref}=\omega t + \phi_0$, but in \eqref{eq:ellipsePhase} the reference angle depends on the amplitudes of the components, so that its rate of rotation is not constant with time. Figure~\ref{fig:RotationsOppositeVsRef} shows the orientation angles of each component vector and of the sum vector at three different snapshots. For instance, from Fig.~\ref{fig:RotationsOppositeVsRef}(a) to Fig.~\ref{fig:RotationsOppositeVsRef}(b), the orientation angle of the sum (green) increases at a slower rate than the orientation angle of the CCW component (blue), resulting in an increase of the angle ($\theta_1 - \theta\subrm{ref}$) between them. From Fig.~\ref{fig:RotationsOppositeVsRef}(b) to Fig.~\ref{fig:RotationsOppositeVsRef}(c), it increases at a higher rate and the angle between them becomes 0. Visualization~2 shows the same data as an animation, in order to clearly show $\theta_3$ changes at different rates along each cycle, exactly as \eqref{eq:ellipsePhase} suggests.

\begin{figure*}[t]
    \centering
    \includegraphics[width=1.0\linewidth]{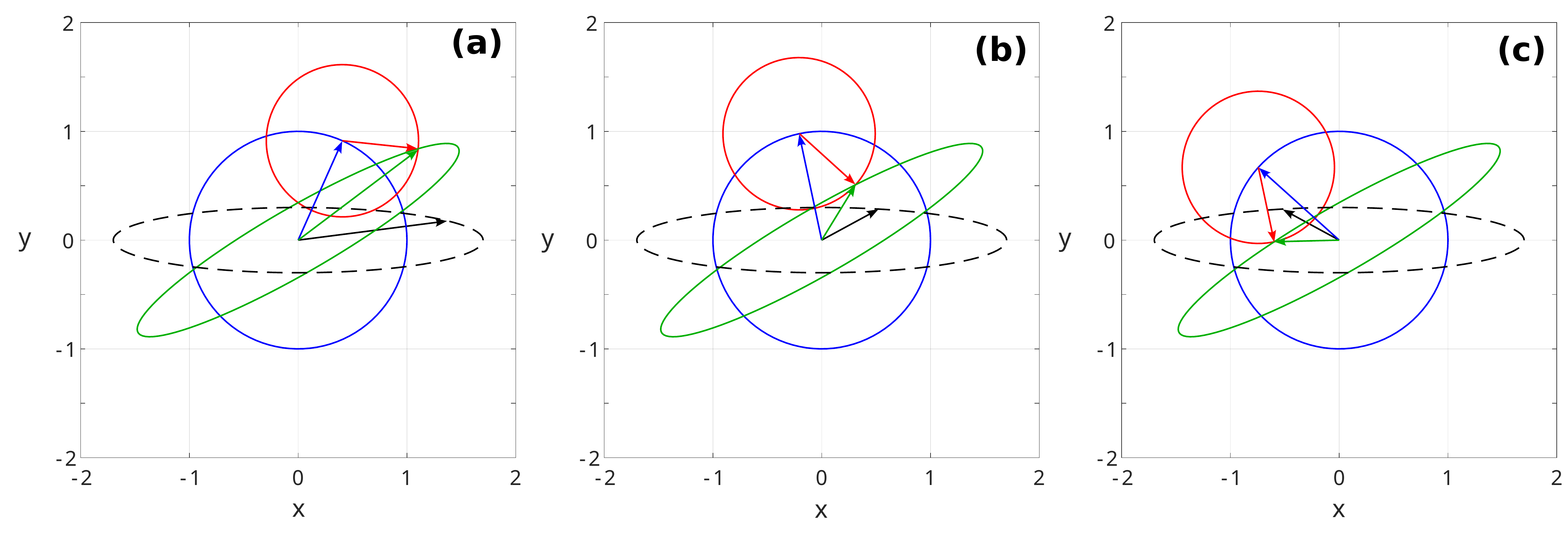} 
    \caption{Addition of two counter-rotating vectors for $A_1 = 1$, $A_2 = 0.7$, and $\delta = 60\degree$ at three time-snapshots: (a) $\omega t = 36\degree$, (b) $\omega t = 72\degree$, (c) $\omega t = 108\degree$. The black dashed ellipse represents the ellipse traced when $\delta = 0$. Visualization~3 shows an animation of this model for $0 \leq \omega t \leq 4\pi$.}
    \label{fig:RotationsOppositeSenseDelayedComponents}
\end{figure*}

In the general case where $\delta \neq 0$, the orientation angle of the sum corresponds to \eqref{eq:t+delta/2}.
Using the co-rotating angle $\theta\subrm{ref}$ as the reference axis allows us to drop $\omega t$ from \eqref{eq:t+delta/2}, equivalent to having an initial phase opposite to the time evolution of the components, $\phi_0=-\omega t$. In that case we arrive at the simple expression 
\begin{equation}
    \gamma = \delta/2,
\end{equation}
which is the geometric phase of the addition of opposite sense rotations. Figure~\ref{fig:RotationsOppositeSenseDelayedComponents} shows three snapshots of the case where $A_1 = 1$, $A_2 = 0.7$, $\delta = 60\degree$, at time-snapshot $\omega t = 18\degree$, 36\degree, \& 54\degree, with the path of the sum vector $\vv{r}_3$ drawn as a green ellipse. That is, the nonzero phase difference $\delta$ between the component vectors causes the sum vector to trace an ellipse that is rotated with respect to the reference angle (the black arrow, given by the angle that the sum vector would have if $\delta = 0$). The angle of the major axis of the green ellipse in \figref{fig:RotationsOppositeSenseDelayedComponents} relative to the major axis of the dashed black ellipse is equal to $\gamma$.

Since rotating vectors can represent a circular polarization basis for the case of optical waves, this result shows that the same phenomenon of superposition-induced shift of the peak location causes a similar phase shift for the case of combining two circularly polarized waves as it does when combining two linearly polarized optical waves. This case of geometric phase produced by superposing two circular polarization states is often labelled ``spin-redirection phase'' in optics, or sometimes as Rytov-Vladimirski-Berry phase.

Equation~\ref{eq:t+delta/2} shows how a rotation of the major axis of an elliptical polarization state is equivalent to introducing a phase difference $\delta$ between the circular polarization components of an optical wave. Therefore, in this basis, a geometric phase can be considered equivalent to a rotation of the polarization state azimuth. This is confirmed in \figref{fig:RotationsOppositeSenseDelayedComponents}, where we see that going from subfigures (a) to (b) and (c), the geometric phase remains constant at $\gamma = \delta/2$: the angle between the sum vector and the reference vector (green and black arrows, respectively) remains constant despite the various changes in amplitude and phase of the two input vectors. Visualization~3 shows an animation of this model over two full rotations $0 \leq \omega t < 4 \pi$.

\section{Three-dimensional propagation paths that induce geometric phase shifts}\label{sec:3D_paths}

The three-mirror system shown in \figref{fig:three_mirrors} is a common model for analyzing geometric phase shifts produced by changes in the propagation direction.\cite{Galvez1999,Nityananda2014,Chipman2019} In this system, the light is initially propagating along the $z$-axis, reflects from the first fold mirror to a new propagation direction along $y$, a second fold mirror to a new propagation direction along $-x$, and a third fold mirror that returns the propagation direction to $z$. If we are careful to analyze the behavior of the local coordinate system at each reflection, as indicated by the red and blue arrows shown in the figure, we can see that each p-polarized reflection flips the p-polarization axis. After the three reflections, the optical wave has experienced three such flips, but because each flip is performed along a different axis, the non-commutativity of 3D reflections produces a rotation of the local coordinate system. In the case of \figref{fig:three_mirrors}, we can see that the input coordinate system has been flipped (since an odd number of reflections produces a parity change) and also rotated. If we consider the flip to be $x \to -x$, then the final rotation angle shown is $+45\degree$ with respect to the global coordinate system. If we consider the flip to be about a different axis, then the final rotation angle will be different: i.e., considering it as a $y \to -y$ flip would produce a $-45\degree$ rotation.

Representing this three-reflection system on the sphere of directions gives the spherical triangle drawn in \figref{fig:three_mirrors}(c). The light enters the system propagating along the $+z$-axis, represented by the white dot at the pole of the sphere. After reflecting from the first fold mirror, the propagation path points along the $+y$-direction, given as another white dot on the sphere. Subsequent reflections cause the propagation to point along $-x$ and then back to $+z$. Although the physical path of the light only consists of discrete points rather than continuous paths on the sphere, the rule~\cite{Courtial1999} for calculating the geometric phase is to connect each successive point with the shortest geodesic between them. Once all of the combined geodesic paths trace a closed region of the sphere, we can calculate the subtended solid angle. In the case shown, the region is an octant of the sphere, so that solid angle is $\Omega = \pi / 2$. This corresponds to a geometric phase of $\gamma = -\Omega = -\pi / 2$, which is also the $-90\degree$ rotation angle of the coordinate system (after implementing the parity change), as illustrated in \figref{fig:three_mirrors}. Note that this calculation differs by a factor of two from the solid angle rule used for obtaining the geometric phase from the Poincar{\'e} sphere, where $\gamma\subrm{PB} = -\Omega\subrm{Poincar\'e} / 2$, due to the doubling of physical angles in the definition of the Poincar{\'e} sphere.

The case of a circular polarized wave is shown in \figref{fig:three_mirrors}(b), where the clock angle  at which the arrow is drawn is used to indicate the phase of the wave. Using red and blue to represent right- and left-circularly polarized light, the figure shows how the sense of rotation flips after each reflection. In the end, the situation is the same as that for the Cartesian basis used in \figref{fig:three_mirrors}(a): after implementing the parity change, the remaining effect is a $-90\degree$ rotation.

\begin{center}
   \includegraphics[width=0.95\linewidth]{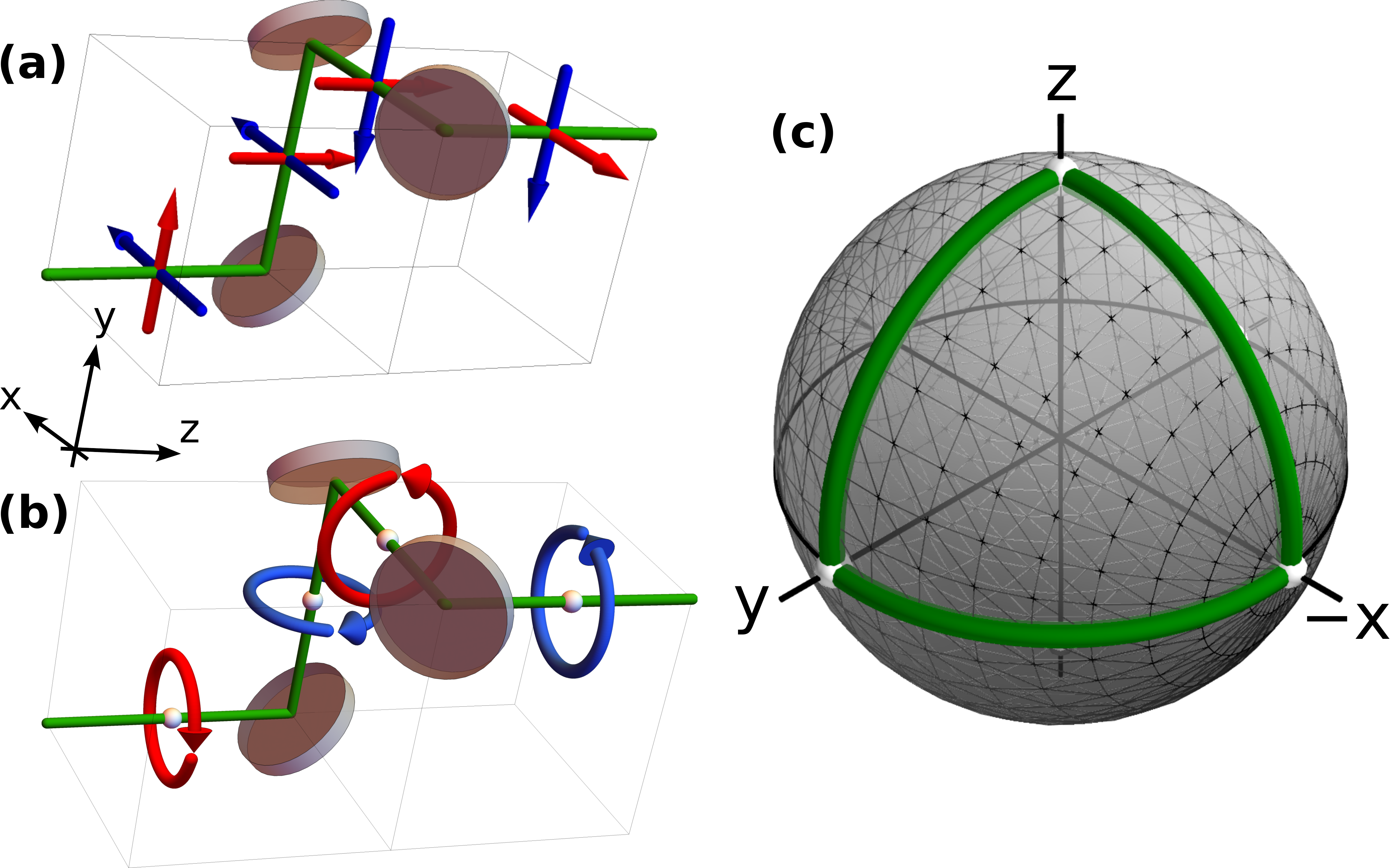}
   \figcaption{(a) A three-fold-mirror system showing the optical axis (green) and polarization basis vectors (blue and red arrows). (b) The same system as (a) but represented using an input right-circularly polarized wave. Right-circularly polarized light is drawn in red, while left-circularly polarized is drawn in blue. (c) The propagation path represented on the sphere of directions.}
   \label{fig:three_mirrors}
\end{center}

\begin{center}
   \includegraphics[width=0.95\linewidth]{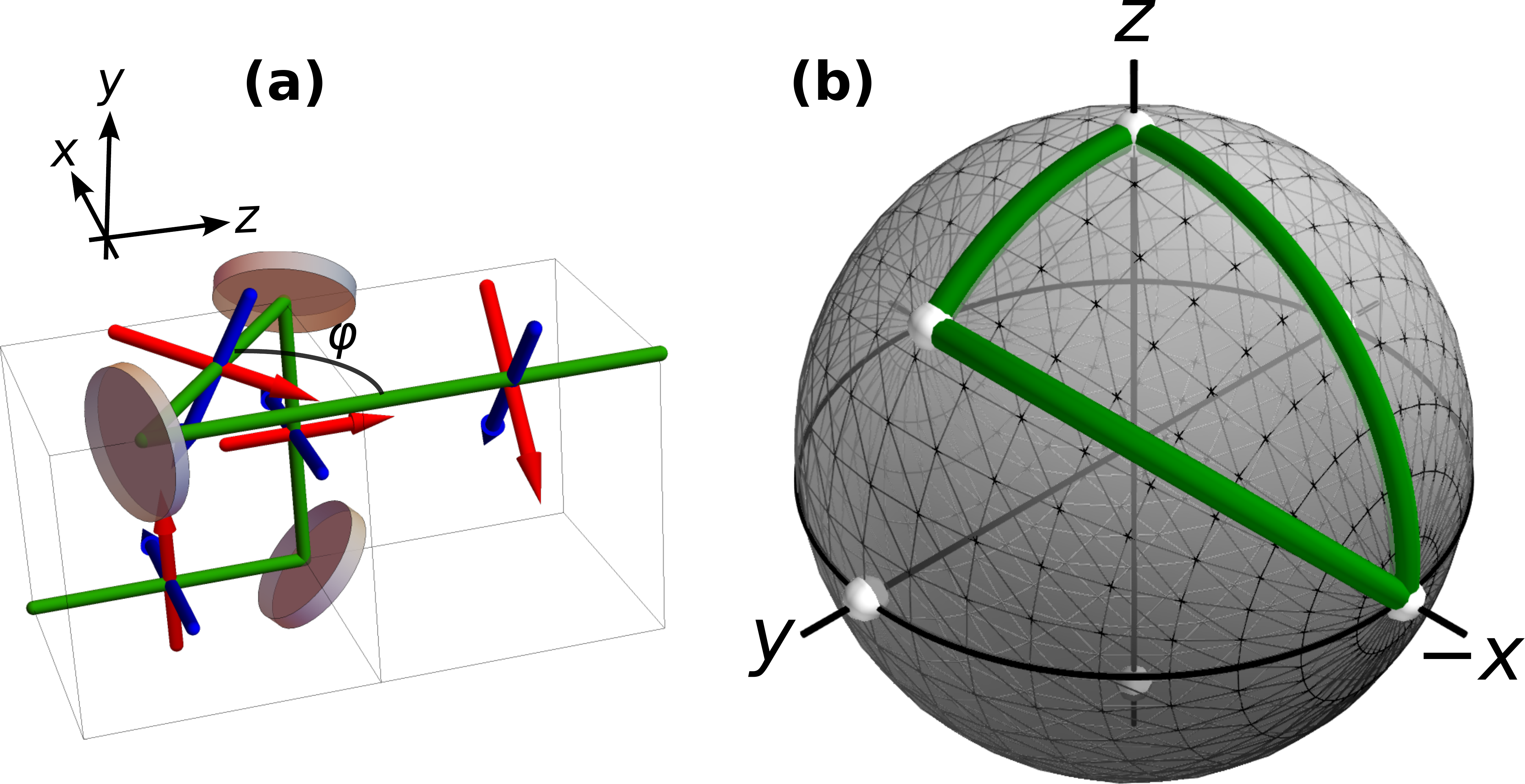}
   \figcaption{(a) A three-fold-mirror system showing the optical axis (green) and polarization basis vectors (blue and red arrows) for case of $\phi = 45\degree$. (b) The propagation path represented on the sphere of directions.}
   \label{fig:skew_mirrors}
\end{center}

Figure~\ref{fig:skew_mirrors} shows a path where the second fold mirror has been rotated by an additional angle of $\tfrac{\pi}{2} - \phi$ in the $x$-$z$ plane. This skew-angle is then corrected by a corresponding rotation of the third fold mirror, but we also see that the changes have produced a $\phi$ rotation of the local coordinate system with respect to the global coordinates. In the extreme case of $\phi = 0$, the rotation angle (and geometric phase) becomes zero. Figure~\ref{fig:skew_mirrors} shows the corresponding path on the sphere of directions for the case of $\phi = 45\degree$. The enclosed solid angle is $\pi / 4$ steradians, so that the geometric phase becomes $\gamma = -\pi/4$ radians, in agreement with the coordinate arrows drawn.

The representation of coordinate systems, and equivalently of geometric phase, shown in Figures~\ref{fig:three_mirrors}(a,b)\,\&\,\ref{fig:skew_mirrors}(a) shows a method of analyzing a system in which propagation phase is imagined to not exist. Without propagation phase, the phase-indicating arrows remain unaffected as they pass through the system, until altered by geometric effects --- a useful technique for visualizing how geometric phase shifts can be separated from propagation phase shifts. This illustrates that there is a concrete alternative to the existing abstract methods of calculating the geometric phase.

Another common model for analyzing geometric phase shifts produced by changes in the propagation direction is that of a helical path. This can be produced by light propagating along an optical fiber that has been wound around a cylinder, so that the fiber axis is a helix. A circular helix $\vv{r} (\phi)$ of radius $R$ and pitch $P$ is a space curve given by $\vv{r} = (x,y,z)$ where
\begin{equation*}
   x = R \cos \phi, \quad y = R \sin \phi, \quad z = P \phi / (2 \pi) \, .
\end{equation*}
and $\phi$ is the curve parameter representing the helix's winding phase. For a helix containing five loops, $\phi$ would therefore extend from 0 to $10 \pi$. The curvature and torsion for a helix with the above definitions are
\begin{equation*}
   \kappa = R / C^2 \, , \qquad \tau = P / C^2 \, .   
\end{equation*}
for $C = \sqrt{R^2 + (P / 2 \pi)^2}$. The pitch angle of the helix --- the angle between the $z$-axis and the unit tangent vector $\vv{T}$ --- is given by $\xi = \arctan (2 \pi R/P)$.

\begin{figure*}[t]
    \centering
    \includegraphics[width=0.95\linewidth]{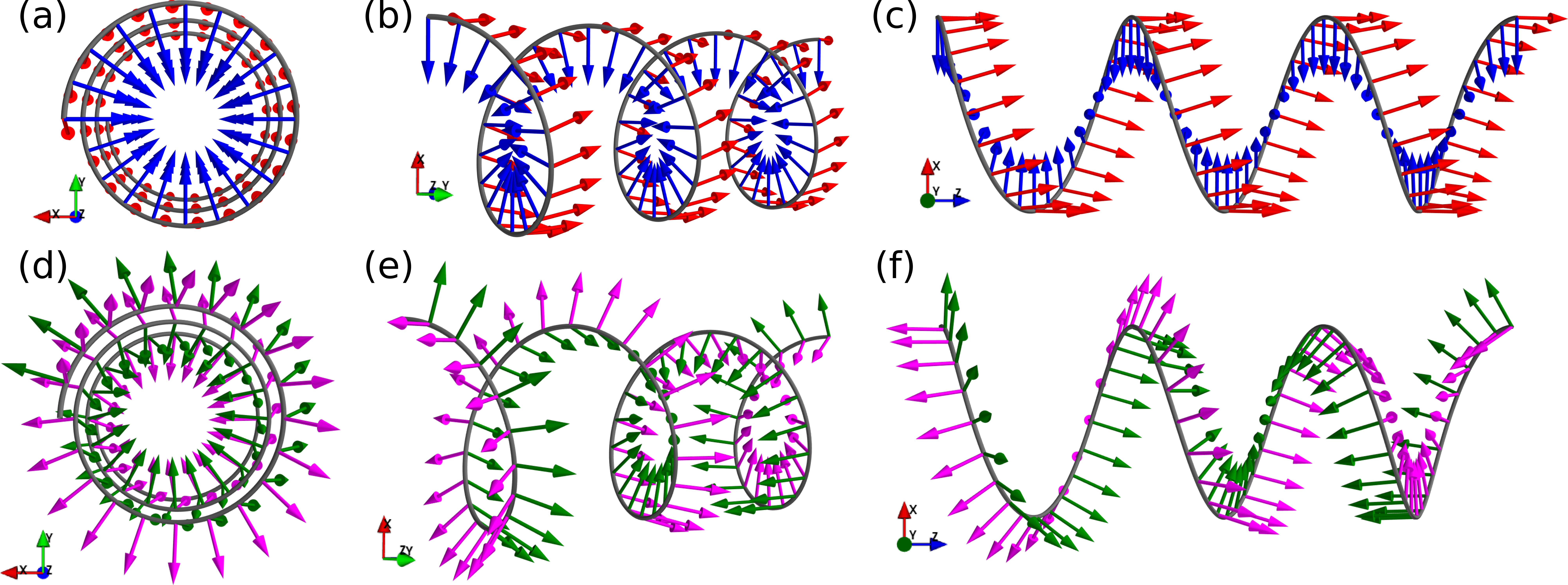}
    \caption{A helix with parameters $R = 1$, $P = 2$, and $N = 3$, labelled with (top row) Frenet-Serret frame coordinates, and (bottom row) rotation-minimizing (parallel transport) frame coordinates. In each case, the helix tangent vectors are not shown. The left, middle, and right columns show views of the helix along the helix axis ($z$), at an oblique angle, and from the side.}
    \label{fig:both_helices}
\end{figure*}

In order to represent an electromagnetic field propagating along a helix, we need a frame that determines the coordinate system. A convenient choice is the Frenet-Serret frame, defined by a triplet of unit-length basis vectors $(\vv{T},\vv{N},\vv{B})$ --- the tangent, normal, and binormal. At each point along the helix, the unit tangent vector is given by
\begin{equation*}
   \vv{T} (\phi) = (d \vv{r} / d \phi) / \norm{d \vv{r}' / d \phi} \, ,
\end{equation*}
and the normal \& binormal vectors by
\begin{equation*}
   \vv{N} (\phi) = (d \vv{T} / d \phi) / \norm{d \vv{T} / d \phi} \, , \quad \vv{B} (\phi) = \vv{T} (\phi) \times \vv{N} (\phi) \, .
\end{equation*}
For any given point on the helix, the normal vector $\vv{N}$ is perpendicular to the tangent vector $\vv{T}$ and points towards the helix's center of curvature; the binormal vector $\vv{B}$ is perpendicular to both the $\vv{T}$ and $\vv{N}$, completing the orthonormal basis. Figure~\ref{fig:both_helices} shows the Frenet-Serret frame vectors along the helix, where normal and bi-normal vectors are drawn in blue and red, respectively. (The tangent vector is not shown.) One characteristic of the Frenet-Serret frame (visible in \figref{fig:both_helices}(a)) is that after a full turn of the helix, the coordinate system returns to its same orientation, with only a shift of its origin.

The Frenet-Serret (FS) frame, however, is not the only available choice. If we maintain the tangent vector as one element of the frame, then we have a rotational degree of freedom over where to place the two coordinates transverse to $\vv{T}$. Another choice is the rotation minimizing (RM) frame, sometimes referred to as the ``Tang frame''~\cite{Tang1970} obtained by rotating the transverse coordinates relative to the Frenet-Serret frame by an angle $\theta$ given by~\cite{Bishop1975,Ross1984,Guggenheimer1989}
\begin{equation}\label{eq:theta}
   \theta = \int_0^{\phi} \tau (\phi') \, d \phi' \, .
\end{equation}
The advantage of doing this, as implied in the name, is that the RM frame minimizes the overall twist $\theta$ that the coordinate system is subjected to with respect to the tangent vector. In particular, the rotation-minimizing frame also happens to be the frame in which the electric field vector is parallel transported when represented on the sphere of directions. For this reason, it is also referred to as the ``parallel transport frame''.

The FS frame rotates about the tangent vector $\vv{T}$ at a rate of $\tau$ per unit path length, while the RM frame counters this rotation by the opposite amount. But while the RM frame does not rotate with respect to the tangent vector, there is still a rotation with respect to the global coordinate system. Since the FS frame repeats itself after each winding, and therefore re-aligns itself with the global frame after each loop, this counter-rotation of the RM frame also means that we can generally expect the RM frame at the end of the helix to be rotated with respect to the global frame. This is exactly what previous research has found, with the amount of rotation equal to $\theta$ (\eqref{eq:theta}). One can also use the solid angle approach to calculate the rotation, since the rotation angle corresponds exactly to the geometric phase shift induced by the coordinate rotation. Looking at the propagation path on the sphere of directions, we find that the propagation vector traces out a circle on the surface of the sphere of directions (\figref{fig:helix_path_on_bloch_sphere}). Since the helix pitch angle $\xi$ is also the zenith angle on the sphere, we can easily calculate the geometric phase of the path shown in \figref{fig:helix_path_on_bloch_sphere} using~\cite{Chiao1986,Morinaga2007}
\begin{equation}
   \Omega = \phi \big[ 1 - \cos(\xi) \big] \, ,
\end{equation} 
for winding phase $\phi$ (increases by $2 \pi$ at each loop), and $\gamma = -\Omega$.

\begin{center}
    \includegraphics[width=0.5\linewidth]{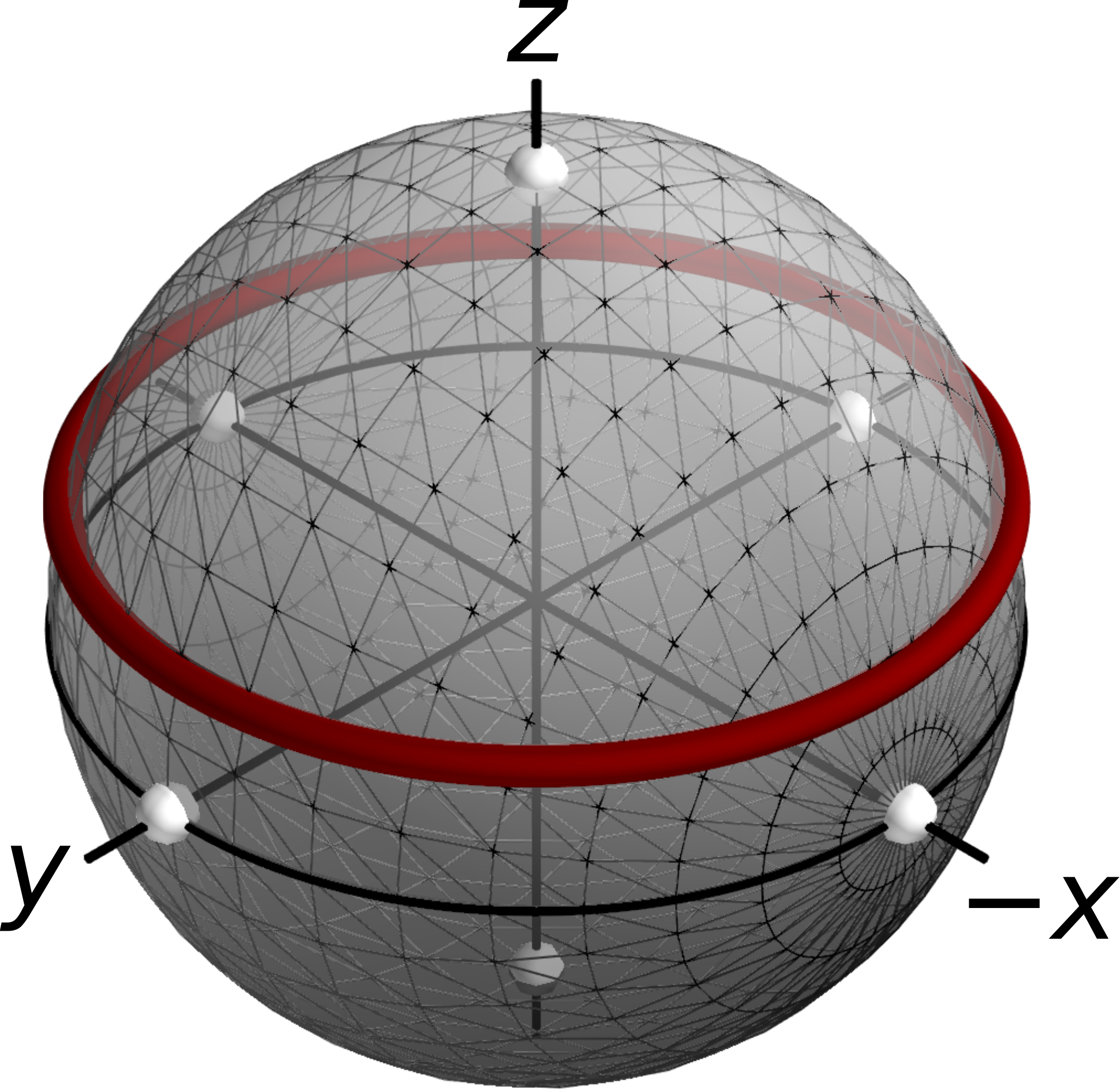}
    \figcaption{The path on the sphere of directions for the propagation along a helical path. Propagation along each winding of the helix produces a full turn of the circle around the sphere.}
    \label{fig:helix_path_on_bloch_sphere}
\end{center}


For the helix shown in \figref{fig:both_helices}, $R = 1$, $P = 2$, and $N = 3$ so that the zenith angle is $\xi = 72.34\degree$, and the subtended solid angle of the path shown in \figref{fig:helix_path_on_bloch_sphere} is $\Omega = -250.81\degree \times N$. The helix has $N = 3$ windings, and so the total accumulated solid angle becomes $-752.43\degree$. Wrapping this within a $\pm 180\degree$ range gives $\overline{\Omega} = -32.43\degree$ (the overbar indicates a wrapped result). The wrapping is necessary if we want to compare the result with alternative methods of calculation. For example, if we trace the angle of the helix's RM frame with respect to the global frame, we obtain a total rotation angle (at the end of the helix) of $\theta = -392.41\degree$. When wrapped, this gives $\bar{\theta} = -32.41\degree$, in agreement to within the fourth digit of precision of the solid angle result.

\section{Conclusion}

We have seen that the geometric phase derived from the fold mirror and helical fiber examples is driven entirely by changes between local and global coordinate systems. In the circular polarization basis, a rotation of the local coordinate system with respect to a global reference system produces a phase shift because the orientation of the polarization vector is coupled to the definition of phase in that basis. In a linear polarization basis, a phase shift is coupled to modulation in the amplitude of the polarization vector. Thus PB phase, which is induced by changes in the wave's polarization state, and RVB phase, which is induced by changes in the local versus global coordinate system, are two aspects of the phase viewed from two different perspectives.

When the polarization state of a wave changes at the same time that the local coordinate system changes with respect to the reference global coordinate system, both PB and RVB phase will play a role in the result. We have shown that, in the superposition model of geometric phase, incorporating both changes in the calculation is straightforward: if we know the polarization state representation on the local basis, we project this state onto the global basis and recalculate $\gamma$ using \eqref{eq:tan2L} on the global basis, using~\cite{garza2023wave}
\begin{equation}\label{eq:tan2L}
   \tan (2 \gamma) = \frac{A_1^2 \sin (2 \phi_1) + A_2^2 \sin (2 \phi_2)}{A_1^2 \cos (2 \phi_1) + A_2^2 \cos (2 \phi_2)} \, ,
\end{equation}

In other words, geometric phase is exclusively a property of the superposition of waves. Changes to the polarization state alter the amplitudes and phases of the two component waves being superposed. Changes to the coordinate system likewise modify the projection of the polarization state onto the new basis vectors. A shift in the peak location is the direct result of changing the definition of what the location of a peak means in the new coordinate system.

This straightforward interpretation of the PB and RVB phases allows us to treat both phenomena under one framework, and without any of the abstractions that alternative methods require. While these alternative methods --- differential geometry, and the subtended solid angle on a unit sphere --- provide correct results, these abstractions easily misdirect our attention from the basic underlying physics. For example, the procedure outlined above shows that if we know the input polarization state \& coordinate basis, and the output polarization state \& coordinate basis, the path between the two is no longer relevant for calculating the geometric phase of the resulting wave, as long as we are careful to keep track of the phase reference along the way.\cite{Hagen2024k} What can cause confusion here is that changes in the coordinate system often cause implicit shifts in the phase reference, and this needs to be taken into account in order to get the correct value for changes in the wave phase.


\begin{thebibliography}{25}
   \providecommand{\enquote}[1]{``#1''}
   \providecommand{\url}[1]{{\tt #1}}
   \providecommand{\href}[2]{#2}

   \bibitem[1]{Gutierrez-Vega2011}
   J. C. Gutiérrez-Vega, \enquote{\href{https://doi.org/10.1364/OL.36.001143}{Pancharatnam-{B}erry phase of optical systems},} Opt. Lett. \textbf{36},{ }1143--1145 (2011).

   \bibitem[2]{Garza-Soto2023a}
   L. Garza-Soto, N. Hagen, and D. Lopez-Mago, \enquote{\href{https://doi.org/10.1364/JOSAA.485485}{Deciphering {P}ancharatnam's discovery of geometric phase: retrospective},} J. Opt. Soc. Am. A \textbf{40},{ }925--931 (2023).

   \bibitem[3]{Rytov1938}
   S. M. Rytov, \enquote{Sur la transition de l'optique ondulatoire {\`a} l'optique g{\'e}om{\'e}trique [{O}n the transition from wave to geometrical optics],} Dokl. Akad. Nauk. SSSR \textbf{28},{ }263--266 (1938).

   \bibitem[4]{Vladimirski1941}
   V. V. Vladimirski, \enquote{{\"U}ber die drehung der polarisationsebene im gekr{\'u}mmten lightstrahl [{T}he rotation of polarization plane for a curved light ray],} Dokl. Akad. Nauk. SSSR \textbf{31},{ }222--225 (1941).

   \bibitem[5]{Berry1987b}
   M. V. Berry, \enquote{Interpreting the anholonomy of coiled light,} Nature \textbf{326},{ }277--278 (1987).

   \bibitem[6]{Tavrov1999}
   A. V. Tavrov, T. Kawabata, Y. Miyamoto, M. Takeda, and V. V. Andreev, \enquote{\href{https://doi.org/10.1364/JOSAA.16.0919 - STUDY THIS}{Method to evaluate the geometrical spin-redirection phase for a nonplanar ray},} J. Opt. Soc. Am. A \textbf{16},{ }577--579 (1999).

   \bibitem[7]{Berry1984}
   M. V. Berry, \enquote{Quantal phase factors accompanying adiabatic changes,} Proc. Roy. Soc. London A \textbf{392},{ }45--54 (1984).

   \bibitem[8]{Aharonov1987}
   Y. Aharonov and J. Anandan, \enquote{\href{https://doi.org/10.1103/PhysRevLett.58.1593}{Phase change during a cyclic quantum evolution},} Phys. Rev. Lett. \textbf{58},{ }1593--1596 (1987).

   \bibitem[9]{Hannay1985}
   J. H. Hannay, \enquote{Angle variable holonomy in adiabatic excursion of an integrable {H}amiltonian,} J. Phys. A \textbf{18},{ }221--230 (1985).

   \bibitem[10]{Berry2011}
   M. V. Berry and P. Shukla, \enquote{Slow manifold and {H}annay angle in the spinning top,} Eur. J. Phys. \textbf{32},{ }115--126 (2011).

   \bibitem[11]{garza2023wave}
   L. Garza-Soto, N. Hagen, D. Lopez-Mago, and Y. Otani, \enquote{Wave description of geometric phase,} J. Opt. Soc. Am. A \textbf{40},{ }388--396 (2023).

   \bibitem[12]{Nityananda2014}
   R. Nityananda and S. Sridhar, \enquote{Light beams with general direction and polarization: global description and geometric phase,} Annals of Physics \textbf{341},{ }117--131 (2014).

   \bibitem[13]{Jiao1989}
   H. Jiao, S. R. Wilkinson, R. Y. Chiao, and H. Nathel, \enquote{Two topological phases in optics by means of a nonplanar {M}ach-{Z}ehnder interferometer,} Phys. Rev. A \textbf{39},{ }3475--3487 (1989).

   \bibitem[14]{Tavrov2000}
   A. V. Tavrov, Y. Miyamoto, T. Kawabata, M. Takeda, and V. A. Andreev, \enquote{\href{https://doi.org/10.1364/JOSAA.17.000154}{Generalized algorithm for the unified analysis and simultaneous evaluation of geometrical spin-redirection phase and {P}ancharatnam phase in a complex interferometric system},} J. Opt. Soc. Am. A \textbf{17},{ }154--161 (2000).

   \bibitem[15]{garza2024differences}
   L. Garza-Soto, N. Hagen, D. Lopez-Mago, and Y. Otani, \enquote{Differences between the geometric phase and propagation phase: clarifying the boundedness problem,} Appl. Opt. \textbf{63},{ }645--653 (2024).

   \bibitem[16]{Galvez1999}
   E. J. Galvez and C. D. Holmes, \enquote{Geometric phase of optical rotators,} J. Opt. Soc. Am. A \textbf{16},{ }1981--1985 (1999).

   \bibitem[17]{Chipman2019}
   R. A. Chipman, W.-S. T. Lam, and G. Young, \textit{Polarized {L}ight and {O}ptical {S}ystems,} Chap.~7 (CRC Press, 2019).

   \bibitem[18]{Courtial1999}
   J. Courtial, \enquote{\href{https://doi.org/10.1016/S0030-4018(99)00473-3}{Wave plates and the {P}ancharatnam phase},} Opt. Comm. \textbf{171},{ }179--183 (1999).

   \bibitem[19]{Tang1970}
   C. H. Tang, \enquote{An orthogonal coordinate system for curved pipes,} IEEE Trans. MTT \textbf{18},{ }69 (1970).

   \bibitem[20]{Bishop1975}
   R. L. Bishop, \enquote{There is more than one way to frame a curve,} Am. Math. Monthly \textbf{82},{ }246--251 (1975).

   \bibitem[21]{Ross1984}
   J. N. Ross, \enquote{The rotation of the polarization in low birefringence monomode optical fibres due to geometric effects,} Opt. Quant. Elect. \textbf{16},{ }455--461 (1984).

   \bibitem[22]{Guggenheimer1989}
   H. Guggenheimer, \enquote{Computing frames along a trajectory,} CAGD \textbf{6},{ }77--78 (1989).

   \bibitem[23]{Chiao1986}
   R. Y. Chiao and Y.-S. Wu, \enquote{\href{https://doi.org/10.1103/PhysRevLett.57.0933}{Manifestations of {B}erry's topological phase for the photon},} Phys. Rev. Lett. \textbf{57},{ }933--936 (1986).

   \bibitem[24]{Morinaga2007}
   A. Morinaga, A. Monma, K. Honda, and M. Kitano, \enquote{\href{https://doi.org/10.1103/PhysRevA.76.052109}{Berry's phase for a noncyclic rotation of light in a helically wound optical fiber},} Phys. Rev. A \textbf{76},{ }052109 (2007).

   \bibitem[25]{Hagen2024k}
   N. Hagen and L. Garza-Soto, \enquote{\href{https://doi.org/10.1364/JOSAA.41.002014}{Evolution of geometric phase and explaining the geodesic rule},} J. Opt. Soc. Am. A \textbf{41},{ }2014--2022 (2024).
\end{thebibliography}

\end{multicols}

\end{document}